\documentstyle[preprint,prl,aps]{revtex}
\begin{document}

\title{Quadratic tranverse anisotropy term due to dislocations in
Mn$_{12}$-Ac crystals directly observed by EPR spectroscopy}
\author{R. Amigo, E. del Barco, $^*$Ll. Casas, $^*$E. Molins, and J. Tejada}
\address{Departamento de F\'\i sica Fundamental, Universitad de Barcelona\\
Diagonal 647, Barcelona, 08028, Spain\\$^*$Instituto de Ciencia de
Materiales de Barcelona (CSIC). 08193 Cerdanyola. Spain}
\author{I. B. Rutel, $^{**}$B. Mommouton, N. Dalal, and J. Brooks}
\address{NHMFL/FSU, CM/T group 1800
E.P.Dirac dr. Tallahassee, FLORIDA 32310\\$^{**}$Institut National
des Sciences Appliquees. Toulouse. France}
\date{11/01/01}
\maketitle

\begin{abstract}
High-Sensitivity Electron Paramagnetic Resonance experiments have
been carried out in fresh and stressed Mn$_{12}$-Acetate single
crystals for frequencies ranging from 40 GHz up to 110 GHz. The
high number of crystal dislocations formed in the stressing
process introduces a $E(S_x^2-S_y^2)$ transverse anisotropy term
in the spin hamiltonian. From the behaviour of the resonant
absorptions on the applied transverse magnetic field we have
obtained an average value for $E$ = 22 mK, corresponding to a
concentration of dislocations per unit cell of $c$ = 10$^{-3}$.
\end{abstract}

\pacs{75.45.+j, 75.50.Tt, 75.60.Lr}

Since J. Friedman et al. found the stepwise magnetic hysteresis on
Mn$_{12}$-Acetate molecular cluster, and interpreted it in terms
of resonant quantum tunneling \cite{Friedman}, a huge number of
experimental measurements have been carried out on this compound
\cite{Hernandez,Thomas,Luis,Perenboom,Hill,Fort,Wernsdorfer,Bokacheva,Kent}(see
also references herein), showing its quantum behaviour under many
different experimental techniques. Mn$_{12}$-Ac is a spin 10
molecular cluster which forms macroscopic single crystals in which
all the molecules are identical and they take the same spacial
orientation respect to the crystallographic axes. The magnetic
structure of the spin $S$ = 10 of each molecule is represented, in
first approximation, by the Hamiltonian
$H=-DS_z^2-g\mu_B\textbf{S}\cdot\textbf{H}$, where $D$ is the
uniaxial anisotropy constant, g is the Lande factor, $\mu_B$ is
the Bohr magneton and $\textbf{H}$ is the applied magnetic field.
In the absence of external field, opposite sign levels of the
2$S$+1 $S_z$ components of the spin $S$ are degenerated in energy
two to two in the both sides of the anisotropy barrier $U$ =
D$S^2$. When a longitudinal magnetic field, $H_z$, is applied, the
spin levels in the direction of the field become more stable and
the system relaxes towards the new equilibrium state. It has been
firmly established experimentally that this relaxation can occur
via quantum tunneling. However, in the absence of transverse
components of external magnetic fields, the uniaxial Hamiltonian
cannot explain the quantum relaxation. Hamiltonian terms not
commuting with $S_z$ are needed in order to explain the observed
quantum phenomena. Due to the tetragonal symmetry of the
Mn$_{12}$-Ac crystals, transverse anisotropy terms which are
quadratic on the spin operator are not allowed. The lowest order
transverse terms are those proportional to $S_+^4+S_-^4$
\cite{Hartmann}, otherwise, these terms imply that quantum
tunneling can not occur at all the resonant fields, $H_z=nD$,
contrary to the experimental observations in Mn$_{12}$-Ac. To
explain the presence of quantum tunneling at any resonant field
there have been several theoretical approaches, including the
transverse component of the magnetic field due to dipolar
intermolecule interactions or hyperfine interaction with the
nuclei spin of the manganese atoms
\cite{Prokof'ev,Garanin2,Chudnovsky2,Prokof'ev2,Wernsdorfer2,Miyashita}.
However, from the magnetic relaxation experiments
\cite{Friedman,Hernandez,Thomas,Luis,Wernsdorfer,Bokacheva,Kent},
it seems to be necessary stronger magnetic fields than dipolar or
hyperfine fields in order to explain the high value of the quantum
relaxation that Mn$_{12}$-Ac exhibits. Recently, Chudnovsky and
Garanin \cite{Chudnovsky,Garanin} have suggested a new theoretical
approach which explains the quantum behaviour of Mn$_{12}$-Ac in
terms of dislocations existing in the crystals. They propose that
crystal dislocations introduce quadratic terms on $S_x$ and $S_y$,
producing tunneling in a lower order of perturbation theory than
the transverse-field. Unpublished experimental results seem to
support this new theory \cite{Sarachik,Hernandez2,Torres,note}. In
this paper we will show a new experimental approach which
indicates that dislocations formed in a strongly stressed single
crystal introduce the $E(S_x^2-S_y^2)$ term suggested by
Chudnovsky and Garanin, where $E$ is the transverse anisotropy
constant.

The Mn$_{12}$-Ac organometallic cluster forms a molecular crystal
of tetragonal symmetry with the lattice parameters $a$ = 1.732 nm
and $c$ = 1.239 nm \cite{Lis}. The unit cell contains two
Mn$_{12}$O$_{12}$ molecules surrounded by four water molecules and
two acetic acid molecules. In the crystallization process point
defects usually appear in a low number along the whole crystal. It
has been shown experimentally that dislocations can be created in
a Mn$_{12}$-Ac single crystal by rapid thermal cycles, in which
the high change of temperature in a low scale of time produces
radial and tangential tensions between the core and the surface of
the crystal \cite{Hernandez2,Torres}. It is easy to visualize that
tensions and pressure forces produced by mechanical distortion of
the crystal may generate the same kind of dislocations in a single
crystal. This is exactly what we have done with a Mn$_{12}$-Ac
single crystal. When a dislocation is created inside the crystal
it has produced a disorder of the molecules in the vicinity of the
dislocation. As the number of dislocations in the crystal grows
the disorder extends over the whole crystal converting it in a
mosaic crystal. The mosaicity is related to the number of
dislocations existing in the crystal and can be determined
experimentally by analyzing the width of the X-ray difraction
peaks \cite{Barabash,Mathieson,Mathieson2}.

In our experiments we have used two single crystals: a) fresh
single crystal and b) strongly stressed single crystal. To distort
the crystal we put it in a glue by one of the extremities. Then a
gradually increasing force was applied to the other extreme,
perpendicular to the longer length, until the the crystal
fractured. Then, the longer resulting part of the crystal was
adequately cleaned and tested by X-ray diffraction to be sure that
it was a single crystal. Both samples were characterized by X-ray
analysis before making the EPR experiments. To obtain a fine check
of the variation of the mosaicity between fresh and stressed
crystals we have used low angle difraction peaks in order to
minimize the widening associated to the lack of monochromaticity
of the $K_{\alpha 1}$ and $K_{\alpha 2}$ X-ray emission lines of
Mo element. We used a four-circle single-crystal X-ray
diffractometer (Enraf-Nonius CAD4, MoKa radiation) in the
characterisation. Reflections ($\pm$ 2,$\pm$ 2,$\pm$ 2) were
studied for both fresh and stressed single crystals. Peak
intensities of the stressed crystal were normalised according to
the measured volume of the studied fragment. Figure 1 shows
$\Delta\omega-\Delta(2\theta)$ plot of (2,2,-2) reflection for the
fresh (left) and stressed crystal (right). An enlargement of the
peak width along the $\omega$ direction is clearly observed
although there is not a significant change along the 2$\theta$
direction (see inset of figure 1). Assuming that the distance
between dislocations is inversely related to the widening of the
reflection peak in the $\omega$ direction
\cite{Hernandez2,Barabash} we can conclude that the stressed
crystal has a larger mosaicity than the fresh one. As the $\omega$
peak-width in the stressed crystal is a factor about 2 higher than
in the fresh crystal, we can conclude that the number of
dislocations increases about one order of magnitude in the
distortion process.

The high frequency  resonance experiments have been carried out
using the AB millimeter wave vector network analyzer (MVNA)
\cite{Abmm}. The base frequency obtained from this source (range 8
- 18 GHz) is multiplied by Q, V and W Schottkey's diodes to obtain
frequency range used in our experiment (37 - 109 GHz). The sample,
a single Mn$_{12}$-Ac crystal, is placed on the bottom of the
cylindrical resonant cavity, halfway between its axis and
perimeter. The applied dc magnetic field is parallel to the cavity
axis and approximately perpendicular to the easy (c) axis of the
crystal. The experiment frequencies are TE$_{0np}$ ($n$, $p=$ 1,
2, 3, ...) which are the resonant frequencies for the cavity used.
Resonance Q-factor varies from 20000 at TE$_{011}$ mode (41.6 GHz)
to a few thousand at higher frequencies. Due to the high
sensitivity at resonance, there is an increase over conventional
EPR of almost 3 order of magnitude, allowing for the detaction of
the absorption peaks suggested theoretically by the
diagonalization of the corresponding spin Hamiltonian.

The spin Hamiltonian used to explain the experimental data
obtained in the last years is
\begin{equation}
{\cal H}=-DS_z^2-BS_z^4+C(S_+^4+S_-^4)-g\mu _B{\bf H}\cdot {\bf S}
\end{equation}
where $D$, $B$, and $C$ parameters have been experimentally
obtained by EPR, Neutron spectroscopy and magnetic relaxation
experiments
\cite{Friedman,Hernandez,Thomas,Luis,Perenboom,Hill,Bokacheva,Kent,Barra,Mukhin,Zhong,Mirebeau}.
{\bf H} is the applied magnetic field. The component of the
applied magnetic field parallel to the easy axis direction, called
longitudinal component, $H_\|$ = $H$ $\sin \theta$, changes the
barrier height between the two classical spin orientations, while
the component of the field on the hard plane, transverse
component, $H_\bot$ = $H$ $\cos \theta$, affects the overlapping
of the respective wave functions, which determines the quantum
splitting of the degenerate spin states. In the absence of
longitudinal field, the quantum splitting of the different $m$
spin levels, $\Delta_m$, and consequently the rate of resonant
tunneling between the spin levels depend on the magnitude of the
transverse component $H_\bot$. The evolution of the spin levels in
the two wells, as a function of the intensity of the transverse
component of the field, can be deduced from the diagonalization of
the spin Hamiltonian of equation (1). In the transmission specta
of our EPR experiments we can detect absorptions peaks
corresponding to the absorption of radiation of frequency $f$ by
the transitions effectuated between the spin levels of the
Hamiltonian with energy difference equal to $hf$. From the field
position of these peaks we can extract the behaviour of the
quantum splittings on the transverse magnetic field. The results
obtained with the fresh crystal are plotted in figure 2 (solid
circles). The experiments have been done at frequencies ranging
from 50 GHz up to 110 GHz and fields up to 9 T, placing the
crystal with the easy $z$-axis perpendicular to the field
direction. We have found the best fitting of our data for $D$ =
555 mK, $B$ = 1.3 mK, and $C$ = 2.2$\times10^{-2}$ mK (solid lines
in figure 2), in good agreement with the values given in
references \cite{Hill,Barra,Mirebeau}. From this figure, it is
clearly observed that the dependence of the quantum splittings
($\Delta_{10}$ first right line, $\Delta_{9}$ second, $\Delta_{8}$
third, and so on) on the transverse field is matching perfectly
with the theoretical calculation. The lines that appear at high
frequencies and round going down and up with the field correspond
to the energy difference between different quantum splittings.
However, there are four points appearing at low frequencies
between ground splitting and first excited splitting lines which
do not have a conventional explanation. These peaks appear also in
the experiments done by Hill et al. \cite{Hill} and the authors
can not give any explanation. The presence of these peaks
constitute an incognita for us too.

In figure 3 we show the EPR absorption spectra recorded at $f$ =
67 GHz for both fresh (A) and stressed (B) Mn$_{12}$-Ac single
crystals. The labelling used in the figure $\alpha_{m,\varphi}$
refers to the $\Delta_m$ splitting absorptions with the field
applied perpendicular to the easy axis with an angle, $\varphi$,
with respect to the $x$ magnetic axis. For a fresh crystal,
represented by the Hamiltonian (1), there is a symmetry between
any direction perpendicular to the easy axis. The peaks observed
for fresh crystal correspond to the resonances with the splittings
$\Delta_9$, $\Delta_8$, $\Delta_7$, and so on, as has been shown
in figure 2. However, for the stressed crystal, each $\alpha_m$
absorption peak appears doubled. This phenomena can be explained
by the addition of a $E(S_x^2-S_y^2)$ term to the Hamiltonian (1),
as Chusdnovsky and Garanin suggest as an effect of dislocations
\cite{Chudnovsky,Garanin}. This term introduces the hardest
anisotropy along the $x$-axis, while $y$-axis remains as a medium
anisotropy axis. For this reason, different directions of the
applied field give different values of each quantum splitting,
$\Delta_{m,\varphi}$. This is the same behaviour observed in the
powder sample of Fe$_8$ molecular clusters
\cite{Wernsdorfer3,delBarco,delBarco2}. The angular dependence of
$\Delta$ ($\varphi $) at a fixed value of the transverse component
of the field is not monotonic. Because of the shape of the
function $\Delta$ ($\varphi $), for a sample with hard axis
oriented at random, that is with not preference for any angle
$\varphi $, there are two values of $\Delta$ for which the density
of states has a peak. These are the values of the splitting
corresponding to $\varphi =$ 0 and $\varphi = \frac \pi 2$ (see
references \cite{delBarco} and \cite{delBarco2}). In the absence
of dissipation, the contribution of each Mn$_{12}$ molecule to the
imaginary part of the susceptibility is proportional to $\delta
(\omega -\frac {\Delta [\varphi ,H_\bot]} \hbar)$. However, the
total imaginary part of the susceptibility is,

\begin{equation}
\chi'' \propto \int_{0}^{\pi} g(\varphi) \delta ( \omega - \frac
{\Delta [\varphi ,H_\bot]} \hbar ) d\varphi \;\; ,
\end{equation}

\noindent where $g(\varphi )$ is the distribution of molecules on
$\varphi$. For a fresh crystal, having no significant number of
dislocations, there is an equivalence between $x$ and $y$-axes as
the Hamiltonian (1) has no preference for any transverse
direction. Due to this, the splitting does not depend on $\phi$
and the amplitude $A$ of the absorption of electromagnetic
radiation must have only one peak corresponding to $\Delta
[H_\bot]$ = h$f$, as it can be seen in figures 2 and 3A. On the
contrary, in a strongly stressed single crystal of Mn$_{12}$ the
dislocations are randomly affecting the magnetic structure of the
molecular clusters, introducing the term $E(S_x^2-S_y^2)$ in
different manner for each molecule \cite{Chudnovsky,Garanin}. That
is, the fact that the effect of the dislocations existing in a
single crystal depends on its direction along the crystal and on
the distance to a given molecule may be considered as a random
generation of a $x$-hard-axis for each molecule. For this reason,
a Mn$_{12}$-Ac single crystal with a large mosaicity can be
aproximated as a powder sample with $x$-axis of the molecules
oriented at random. In this case, Eq. (2) can be rewritten as
\begin{equation}
\chi'' \propto \int_{0}^{\infty} \delta ( \omega - \frac {\Delta
[\varphi ,H_\bot]} \hbar ) \left (\frac {d\Delta} {d\varphi}
\right )^{-1} d\Delta= \left.\left (\frac {d\Delta} {d\varphi}
\right )^{-1}\right |_{\Delta=\hbar \omega} \;\; .
\end{equation}
\noindent Therefore, there are two field values, solutions of the
equations $\Delta [0,H_{\bot 1}]$ = h$f$ and $\Delta [\frac \pi 2
,H_{\bot 2}]$ = h$f$, at which the amplitude of the absorption is
maximal. Due to this, the two doubled peaks of each splitting
observed in the stressed single crystal (figure 3) can be
attributed to the two mean orientations of the splitting $\Delta$
on the angle $\varphi$, $\Delta_{m,0}$ and $\Delta_{m,\pi/2}$. As
the distance between the two doubled peaks basically depends on
the parameter $E$ we can extract a first value of $E\sim$ 20 mK.

A more precise analysis of the effect of dislocations on the spin
Hamiltonian of Mn$_{12}$-Ac can be achieved by studying the
behaviour of the quantum splittings on the transverse magnetic
field. Thus, EPR experiments have been done in the stressed
crystal for many frequencies ranging from 40 Ghz up to 110 GHz. In
figure 4, it has been plotted the field position of the EPR
absorption peaks found at the experiment frequencies $f$ (open
circles). The EPR data have been fitted by using the magnetic
level structure resulting from the diagonalization of the
Hamiltonian (1), adding the term $E(S_x^2-S_y^2)$ attributed to
the effect of the dislocations. The results of our fitting
procedure are shown in figure 4. Black lines correspond to
$\varphi$ = 0 and blue lines correspond to $\varphi$ = $\pi$/2.
The values of the Hamiltonian parameters used in our fitting
procedure are $D$ = 675 mK, $B$ = 0.9 mK, $C$ =
1.8$\times$10$^{-2}$ mK, and $E$ = 22 mK. Comparing these values
with the values extracted from the fitting of the fresh crystal
absorption peaks of figure 2 one can conclude that dislocations
introduce a general variation of the spin Hamiltonian, apart for
the introduction of the transverse anisotropy term. This effect of
dislocations on the magnetic structure of the molecules on the
vicinity is expected in the theoretical model of Chudnovsky and
Garanin \cite{Chudnovsky,Garanin}. As the authors observe,
dislocations may introduce other effects, for example, transverse
magnetic fields. This effect has not been considered in our
analysis because its complexity. However, we can extract
quantitative information of the number of dislocations existing in
the stressed single crystal by analyzing the distribution of the
generated transverse anisotropies. From the theoretical model
\cite{Garanin}, one can extract the distribution of the logarithm
of the transverse anisotropy, $\ln (E/2D)$, as a function of the
concentration of dislocations per unit cell, $c$. This
distribution function has a maximum at a different value of $E/2D$
depending on $c$. If we assume that the observed absorption peaks
correspond to this maximum in the distribution, we can extract an
approximated value of $c$ from the resulting Hamiltonian
parameters from the fit of the experimentally obtained radiation
spectra. We have obtained $\ln (E/2D)$ = -4.1. This corresponds to
a concentration of dislocations per unit cell of $c\sim$
10$^{-3}$, in good agreement with the theoretical estimation
\cite{Garanin}, and with the experimental results
\cite{Sarachik,Hernandez2,Torres} for a similar samples. Using the
comparison X-ray analysis of the mosaicity for both fresh and
stressed crystals of figure 1, one can observe that the number of
dislocations is increased by almost one order of magnitude than
the fresh sample. We conclude that in a fresh crystal the number
of dislocations per unit cell is approximately $c\sim$ 10$^{-4}$.

Through high-sensitivity EPR experiments carried out in a strongly
stressed single crystal of Mn$_{12}$-Acetate we have directly
obtained the magnitude of the transverse anisotropy term
$E(S_x^2-S_y^2)$, with $E$ = 22 mK, associated to dislocations
existing in the crystal. We have estimated the value of the
concentration of dislocations per unit cell in both fresh and
stressed crystals. It may be also possible that the the combined
effect of a heavy X-ray irradiation dose and subsequent thermal
stressing treatment create a large number of new defective sites
and extends the original ones. These sites will necessarily have a
lower symmetry and could lead to new EPR absorption peaks, with an
$E$-term, such as determined here. A clear effect of lattice
defects on magnetization tunneling has been detected recently
\cite{Sarachik,Hernandez2,Torres}. We also note that an EPR
line-broadening effect of naturally present defects in Mn$_{12}$
and Fe$_8$ single crystals has been reported recently \cite{Park}.
Additional investigations are thus needed in order to clearly
understand the origin of the newly found EPR peaks in the present
work.

This work was supported by MEC Grant number PB-96-0169 and EC
Grant number IST-1929. Work at NHMFL/FSU was supported through a
contractual agreement between the NSF through Grant No.
NSF-DMR-95-27035 and the State of Florida.

\pagebreak

{\bf FIGURE CAPTIONS}\\

{\bf Figure 1:} $\Delta\omega$-$\Delta(2\theta)$ plot of both
fresh (left) and stressed (right) Mn$_{12}$-Ac single crystals for
(2,2,-2) reflection. Increase of imperfections is evidenced by
broadening along $\omega$ direction.\\

{\bf Figure 2:} Resonant peaks from fresh single crystal of
Mn$_{12}$-Ac for different frequencies, ranging from 50 GHz up to
110 GHz, as a function of the magnetic field applied
perpendicularly to the magnetic easy axis direction of the
crystal. The solid lines are the fitting result
of the diagonalization of the spin Hamiltonian of eq. (1).\\

{\bf Figure 3:} EPR spectra recorded at 67 GHz for both fresh (A)
and stressed (B) single crystals of Mn$_{12}$-Ac. The experiments
were done at $T$ = 10 K. The peaks are associated to the different
quantum splittings through the next nomenclature: $\Delta_{m,\varphi}$.\\

{\bf Figure 4:} EPR peaks for the stressed single crystal of
Mn$_{12}$-Ac. The data are fitted by adding the term
$E(S_x^2-S_y^2)$ to the Hamiltonian of eq. (1). The fitting lines
represent the field behaviour of the quantum splittings for
$\varphi$ = 0 (black lines) and $\varphi$ = $\pi$/2 (blue lines). \\

\pagebreak


\begin{references}
\bibitem{note} The combination of both X-ray irradiation and
mechanical stressing may be additive in the process of
dislocations and defects creation.
\end{references}
\end{document}